\documentclass[prb,twocolumn,citeautoscript,amsfonts]{revtex4}
\usepackage{dcolumn}
\usepackage{epsfig}
\usepackage{times}

\begin{document}
\title{Magnetic ordering, electronic structure and magnetic anisotropy
energy in the high-spin Mn$_{10}$ single molecule magnet}
\author{Jens Kortus}\email{j.kortus@fkf.mpg.de}
\affiliation{Max-Planck-Institut f{\"u}r Festk{\"o}rperforschung,
Heisenbergstr.\ 1, D-70569 Stuttgart, Germany}
\author{Tunna Baruah} 
\author{N. Bernstein}
\author{Mark R. Pederson}
\affiliation{Center for Computational Materials Science,
Naval Research Laboratory, Washington, DC 20375-5000, USA}
\date{\today}

\begin{abstract}
We report the electronic structure and
magnetic ordering of the single molecule magnet
[Mn$_{10}$O$_{4}$(2,2'-biphenoxide)$_{4}$Br$_{12}$]$^{4-}$ based on
first-principles all-electron density-functional calculations.  We find
that two of the ten core Mn atoms are coupled antiferromagnetically
to the remaining eight, resulting in a ferrimagnetic ground state
with total spin $S=13$. The calculated 
magnetic anisotropy barrier is found to be 9~K in good agreement with
experiment. 
The presence of the Br anions impact the electronic structure 
and therefore the magnetic properties of the 10 Mn atoms. However,
the electric field due to the negative charges has no significant 
effect on the magnetic anisotropy.
\end{abstract}
\maketitle

The interest in magnetic molecular clusters of transition metal
ions has been continuously growing since the observation of magnetic bi-stability 
of a purely molecular origin in the so-called Mn$_{12}$-ac \cite{Mn12},
which shows a magnetic hysteresis cycle below 4~K similar
to that observed for bulk magnetic materials.
The magnetic bi-stability associated with the hysteresis cycle
has created an interest in these clusters for information storage,
although at low temperature quantum effects affect the reversal
of the magnetization, resulting in steps in the hysteresis \cite{Mn12-steps}. 
This phenomenon of quantum tunneling of magnetization 
is governed by the magnetic anisotropy energy (MAE)\cite{vFleck}
barrier which is due to directional dependencies of the 
spin-orbit-coupling operator.

Recently, Pederson and Khanna have developed a method for accounting
for second-order anisotropy energies\cite{ped-so}.  This method
relies on an exact and simple expression for the spin-orbit coupling
operator used in a second-order perturbative treatment to determine the
dependence of the total energy on spin projection. Initial applications
to the uniaxial Mn$_{12}$-ac lead to a density-functional-based
second-order anisotropy energy \cite{ped1} of 55.7~K, in agreement
with the experimentally deduced values \cite{Mertes,barra-Mn12} 
of 54.8(3)~K or 55.6~K.

Because the second-order anisotropy energy scales with the square
of the magnetization it was generally believed that a high-spin
ground state $S$ would be beneficial for a large barrier.  The
[Mn$_{10}$O$_{4}$(2,2'-biphenoxide)$_{4}$Br$_{12}$]$^{4-}$ cluster
has been reported to have a $S=12$ high-spin ground state \cite{Mn10}
but only a small energy barrier of about 7.7~K.  In this work we
investigate the electronic and magnetic properties and the magnetic
anisotropy energy of this high-spin single molecule magnet. The
information obtained here may be useful in the search for single
molecule magnets with a greater magnetic anisotropy.

\begin{figure}
\begin{center}
  \epsfig{file=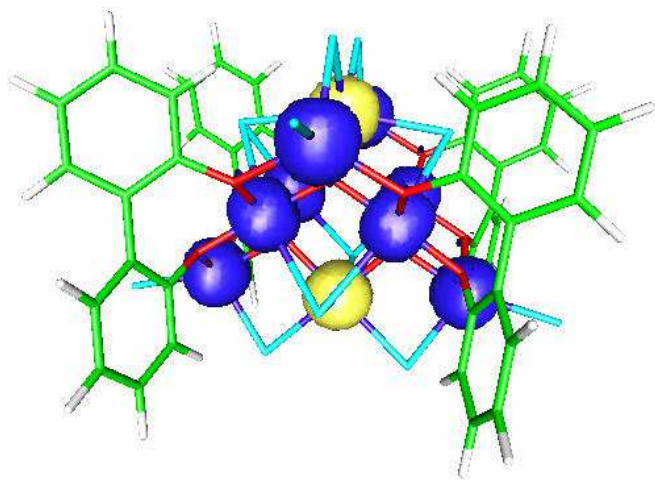,width=\linewidth,clip=true}
\end{center}
\caption{ The tetrahedron like structure of the  ten Mn atoms
and the surrounding organic rings. The Br atoms are not displayed
for clarity.
Density isosurfaces for $0.03$~$e/a_B^3$ for majority (dark)
and minority (light) spins on Mn atoms are shown.
The plot clearly shows that the magnetic
moment is localized at the Mn atoms, and it directly confirms
the antiferromagnetic coupling of two of Mn atoms (large light spheres)
to the remaining Mn atoms
(large dark spheres). 
\label{fig:diff}}
\end{figure}

Fig.~\ref{fig:diff} shows the structure of the 
[Mn$_{10}$O$_{4}$(2,2'-biphenoxide)$_{4}$Br$_{12}$]$^{4-}$ molecular magnet.
The 10 Mn atoms form a tetrahedron like structure with Mn atoms at the
corners and at the middle of the tetrahedron edges. Two of the Mn atoms,
the top and the bottom spheres in Fig.~\ref{fig:diff}, are coupled
antiferromagnetically to the rest of the Mn atoms. The Mn atoms are bridged
by O atoms. The magnetic core is further stabilized by organic rings
that are also connected to the O atoms. 
The negatively charged cluster is compensated by 
[(CH$_3$CH$_2$)$_3$NH]$_2$[Mn(CH$_3$CN)$_4$(H$_2$O)$_2$] in the molecular crystal,
but experimental results suggest that the magnetic anisotropy is due
to the localized valences of the 10 Mn atoms\cite{Mn10}. In order to 
use the higher symmetry and make the problem computationally feasible, 
we carried out our calculation on the negatively
charged [Mn$_{10}$O$_{4}$(2,2'-biphenoxide)$_{4}$Br$_{12}$]$^{4-}$
cluster which contains 114 atoms. The eight symmetry operations
reduce the complete cluster to 18 inequivalent atoms.

The theoretical studies were carried out using a linear combination of
atomic-orbitals molecular-orbital (LCAO-MO) approach within the
density-functional framework \cite{hohenberg,kohn}.
The molecular orbitals were expanded as linear combinations of
Gaussian functions centered at the atomic sites.
The calculations were carried out at the all-electron level
and the multicenter integrals required in the solution
of the Kohn-Sham equation were calculated by integrating numerically
over a mesh of points \cite{pederson1990}.
 
Our density functional-based calculations
were performed with the all-electron Gaussian-orbital based
Naval Research Laboratory Molecular Orbital Library (NRLMOL) program
\cite{pederson1990,jackson1990,briley,porezag1996,pore99,pss2000},
using the Perdew-Burke-Ernzerhof (PBE)
generalized-gradient approximation for the exchange and
correlation functional\cite{pbe}.
NRLMOL combines large Gaussian orbital basis sets, numerically precise
variational integration and an analytic solution of Poisson's equation
to accurately determine the self-consistent potentials, secular
matrix, total energies and Hellmann-Feynman-Pulay forces. 
The exponents for the single Gaussian have been fully optimized
for DFT calculations \cite{pore99}.  The basis set for the Mn$_{10}$
cluster consisted of a total of 3756 contracted orbitals. 
The minimum and maximum exponent of the bare Gaussians, 
the number of bare Gaussians, and the number of contracted 
$s$-, $p$- and $d$-like basis functions are given in Table \ref{tab:basis}
for each atomic species. The contraction coefficients for atomic orbitals
were obtained by performing an SCF-LDA calculation on the spherical
unpolarized atom where the total energy of the atom was converged to within 10 meV.
The basis functions that do not correspond to atomic wavefunction were
constructed from the longest range bare Gaussians in the basis set.
\begin{table}
\caption{The Gaussian basis set used for the calculation. The minimum and maximum
exponent $\alpha$ of the bare Gaussians, 
the number of bare Gaussians, the number of
contracted $s$-, $p$- and $d$-like basis functions.
\label{tab:basis}}
\begin{ruledtabular}
\begin{tabular}{rcccccc}
 & $\alpha_{\text{min}}$ & $\alpha_{\text{max}}$ & N$_{\text{bare}}$ & $s$ & $p$ & $d$ \\
\hline
Br & 0.0781 & 7.9$\times 10^6$ & 21 & 11 & 6 & 4 \\
Mn & 0.0416 & 3.6$\times 10^6$ & 20 & 11 & 5 & 4 \\
 O & 0.1049 & 6.1$\times 10^4$ & 13 &  8 & 4 & 3 \\
 C & 0.0772 & 2.2$\times 10^4$ & 12 &  8 & 4 & 3 \\
 H & 0.0745 & 77.84            &  6 &  5 & 3 & 1 \\ 
\end{tabular}
\end{ruledtabular}
\end{table}

Here we repeat some of the formulas needed for discussion of
the magnetic anisotropy energy.
The same definitions and notation are used as in
Ref.~\onlinecite{ped-so}. In the absence of a magnetic field
the second-order MAE $\Delta_2$ resulting from the spin-orbit coupling,
for an arbitrary symmetry, reduces to
\begin{equation}
\label{eq:1}
\Delta_2=\sum_{\sigma \sigma'}\sum_{ij}
 M^{\sigma \sigma'}_{ij} S^{\sigma \sigma'}_{i} S^{\sigma' \sigma}_{j},
\end{equation}
which is a generalization of Eq.\ (19) of Ref.~\onlinecite{ped-so}.
The matrix elements $S^{\sigma \sigma'}_{i}=\langle \chi^\sigma
|S_i|\chi^{\sigma'}\rangle$ implicitly depend on two angles
($\theta,\beta$) defining the axis of quantization.
The matrix elements $M^{\sigma \sigma'}_{ij}$, which are related to
of the induced orbital moment, are given by
\begin{equation}
\label{eq:M}
 M^{\sigma \sigma'}_{ij}=-\sum_{kl}
\frac{\langle \phi_{l\sigma}|V_i|\phi_{k\sigma'}\rangle
\langle \phi_{k\sigma'}|V_j|\phi_{l\sigma}\rangle}
{\varepsilon_{l\sigma} - \varepsilon_{k\sigma'}},
\end{equation}
where $\phi_{l\sigma}$, $\phi_{k\sigma}$ and $\varepsilon_{l\sigma}$
are, respectively, the occupied,
unoccupied  and the corresponding energies of states. 
$V_i$ is same as defined in  Eq.~(7) of
Ref.~\onlinecite{ped-so} and is related to derivatives of the
Coulomb potential.  
The matrix elements can be evaluated by integrating products of the
Coulomb potential with partial derivatives of the basis functions.
This procedure avoids the time consuming task of calculating the
gradient of the Coulomb potential directly.

In addition to the magnetically interesting complex, the crystal  
also contains single Mn complexes to balance the charges. Using 
high-field EPR spectroscopy Barra {\it et al.} \cite{Mn10} found 
that this
[(CH$_3$CH$_2$)$_3$NH]$_2$[Mn(CH$_3$CN)$_4$(H$_2$O)$_2$]
unit is paramagnetic.  We also find that this unit is paramagnetic,
with the Mn atom in a $+2$ charge state and a spin of $S$=5/2. 
The complex exhibits
easy-plane behavior with an energy well of 0.1~K.  We therefore focus the
remainder of our work on the Mn$_{10}$ unit only.   

Calculations on a $S$=12 high spin state revealed that this
spin state would not be magnetically stable because the Fermi levels 
in the majority and minority spin channel would not be aligned.
The Fermi level misalignment indicated further transport of electrons of
the minority to the majority spin channel.  As a result we obtained a
$S$=13 high spin state as the magnetic ground state instead of the 
$S$=12 state obtained from high-field EPR spectroscopy.
Our result is consistent with experiment since it is difficult
to differentiate experimentally between the two possibilities \cite{roberta}.

A plot of the spin density shown in Fig.~\ref{fig:diff}
clearly confirms the antiferromagnetic coupling of two of the
Mn atoms to the remaining Mn atoms. 
The spin density around the minority spin Mn and the four majority spin Mn at 
the corners of the tetrahedron-like magnetic core
show a nearly spherical spin density
which one would expect for a closed $d^5$ shell.
The other four Mn atoms
which are on the edges of the tetrahedron (large dark spheres) show
a less spherical spin density, indicating another charge state
for these atoms.
In order to analyze the magnetic ordering we calculated the spin density
in spheres around the atoms. This gives a measure of the localized
moment at the atom, but will generally underestimate the exact value.
Using spheres with a radius of 2.23 $a_B$ around the Mn atoms we 
obtained for the three non-equivalent Mn atoms 
(majority spin tetrahedron edge, majority spin tetrahedron vertex, 
and minority spin tetrahedron edge) a local moment 
 of 3.61, 4.33 and -4.33 $\mu_B$, respectively. This result suggests an
ionic picture that the first Mn has an Mn$^{3+}$ ($S=2$) state, 
whereas the other two are Mn$^{2+}$ ($S=5/2$). 
This picture is fully in accord with the spin density plot in Fig.~\ref{fig:diff}.
Due to the symmetry of the cluster, 
the two types of majority spin Mn atoms have a multiplicity of 4 whereas
the minority spin Mn atom has a multiplicity of 2, resulting in the previously
mentioned $S=4\times 2 + 4 \times 5/2 - 2\times 5/2 =13$
magnetic ground state.

\begin{figure}
\begin{center}
  \epsfig{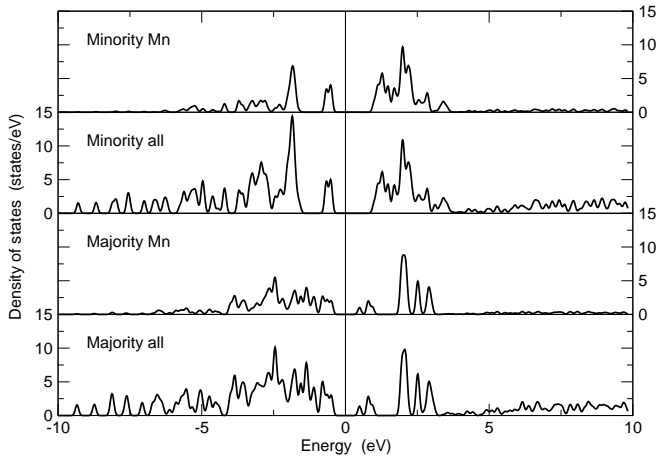}
\end{center}
\caption{
Electronic density of states (DOS) broadened by 0.54 eV of
Mn$_{10}$ in the spin $S=13$ configuration. For each spin the
total DOS and the projected DOS of all Mn($3d$)
are presented. The vertical line indicates the Fermi level. 
\label{fig:dos}}
\end{figure}
The electronic density of states (DOS)
for the majority and minority spin channels is shown in Fig.~\ref{fig:dos}.
For each spin, the DOS is further decomposed into the $3d$ contributions 
of all Mn atoms. It is evident from the plot that the states around
the Fermi level are clearly connected with $3d$ states of the Mn atoms.
This result also agrees well with the experimental picture that the 
states near the Fermi level are well localized and do not show strong
hybridization with other atoms, although we find some 
oxygen, bromine and nitrogen contributions for the 
occupied states.

Starting from the experimental geometry \cite{roberta} we carried out about
30 steps of a conjugate-gradient algorithm using the 
Hellmann-Feynman-Pulay forces 
for optimization of the geometry.
For each new geometry we calculated the complete Hamiltonian of the
magnetic anisotropy and the second-order contribution to magnetic
anisotropy barrier $D S_z^2$. 
In accord with experimental data, we find that the Mn$_{10}$ 
single molecule magnet is an easy-axis system.
The barrier showed no strong dependence on geometry
varying between 8.8~K and 10.4~K with a value of 
9.5~K for the lowest energy geometry. 
Expressing the barrier in the form of a simple spin Hamiltonian
$H= D S_z^2$ we obtained a value of $D=-0.056$~K.
For the calculation of the spin orbit matrix elements, we included all
valence electrons and all unoccupied states in an energy
window of 13.6 eV above the highest occupied state.
The difference between the second-order
treatment and exact diagonalization of the Hamiltonian including the
spin-orbit matrix elements, which includes some higher 
order effects too, was of the order of 0.1~K. 

Eq.~\ref{eq:M} shows that the barrier is related to matrix elements
between occupied and unoccupied orbitals in the majority and minority
spin channels.
In order to give a deeper insight into which states are forming the 
barrier,
we analyze these contributions in more detail.
First, we focus on the contributions of the different spin channels.
Table \ref{tab:mae} summarizes the result in form of the $D$ parameter
allowing only a given spin channel, for example including 
only matrix elements $M_{ij}^{\sigma\sigma'}$ between occupied majority
states and unoccupied minority states, in the calculation of the 
barrier.
All matrix elements from the occupied majority electrons prefer an
easy-axis system, whereas the matrix elements from the occupied minority
spin channel would result in an easy-plane system. Only due to the
larger values of the contributions of the occupied majority spin channel
the system ends up as an easy-axis system. This destructive interference
between the different spin channels seems to be the reason for the
small barrier compared with Mn$_{12}$. In Mn$_{12}$ constructive interference
between the different spin channels was observed.
\begin{table}
\begin{ruledtabular}
\begin{tabular}{cccc}
occupied  & unoccupied  &  $D$  (K)  &  \text{$D S_z^2$ (K)}  \\ \hline
majority  & majority    & -0.039     &    -6.6  \\
majority  & minority    & -0.106     &   -17.9  \\
minority  & majority    &  0.034     &     5.7  \\
minority  & minority    &  0.055     &     9.3 \\
\end{tabular}
\caption{The contributions of the different spin channels
(see Eq.\ \protect\ref{eq:M}) to
the magnetic anisotropy parameter $D$ and the magnetic anisotropy
energy $D S_z^2$.
\label{tab:mae}
}
\end{ruledtabular}
\end{table}
\begin{figure*}
\epsfig{file=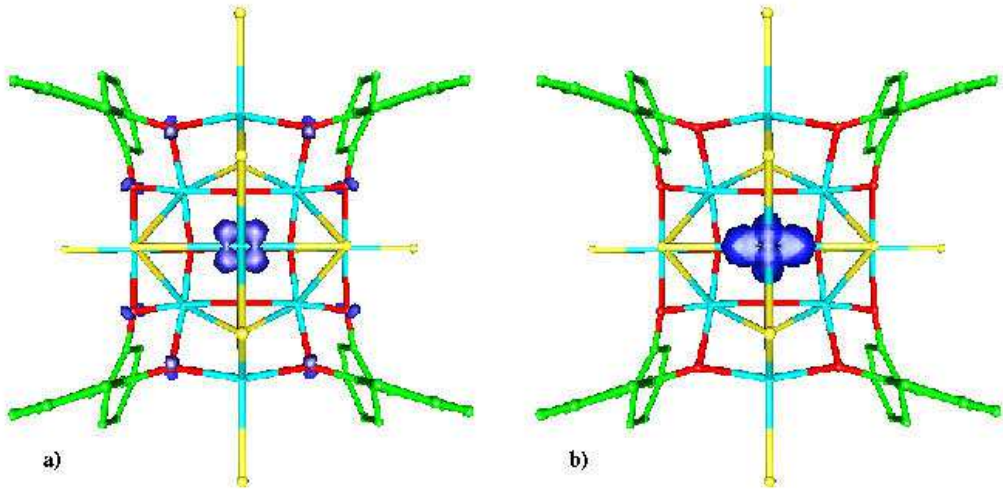,width=7cm,angle=90,clip=true}
\caption{Isolines at 0.005 $e/a_B^3$ of the square of the wavefunctions
(occupied majority state (a) and unoccupied minority state (b) )
which contribute most to the matrix elements $M_{ij}^{\sigma\sigma'}$.
The view is from top with respect to the earlier figures.
It is clearly visible that the matrix element connects majority and
minority $d$-states at the same Mn atom. 
\label{fig:wave}}
\end{figure*}

Besides the spin channel contribution, we can analyze 
which electronic states contribute most
to the matrix elements $M_{ij}^{\sigma\sigma'}$. In Fig.~\ref{fig:wave}
we display plots of the square of the wavefunctions of the occupied
majority state and the unoccupied minority state which contribute to the 
matrix element $M_{ij}^{\sigma\sigma'}$ with the largest absolute value.
The view is from the top with respect to Fig.~\ref{fig:diff}.
It is clearly visible that the states of interest are $d$-states
localized at the same Mn atom. In this case, the states are localized
at the minority spin Mn atoms (light spheres in Fig.~\ref{fig:diff}).
In order to emphasize the $d$ character of the wavefunctions,
we have chosen the top view, although the wavefunctions of the 
other minority Mn atom are just below the top ones and are not visible.

While Mn is the only magnetically active species in the complex,
the remaining atoms affect the magnetic properties of the molecule.
In particular the electric field of the twelve Br$^-$ ions can
affect the MAE through its effects on the electronic structure and
on the spin-orbit coupling.  Effects of the Br ions on the electronic
structure are of a chemical nature, and a detailed analysis is beyond
the scope of this work.  However, direct effects on the spin-orbit
coupling energy could raise the possibility that even small variations
in the positions of these ions could affect the magnetic properties
of the molecule.

To measure the effects of the Br$^-$ ions on the electronic structure
we redid the calculations with various subsets of the Br$^-$ ions.
For each removed Br we also removed an extra electron, keeping
the remaining molecule isoelectronic with the original complex.
For these systems with either zero, four, or eight Br atoms, we
observed a range of behaviors.  In some cases the electronic structure
near the Fermi level was similar to the original molecule, although
it never showed truly rigid-band-like behavior.  In other cases,
however, the electronic structure changed significantly, sometimes
completely closing the HOMO-LUMO gap.  Associated with these electronic
structure changes were large changes in the MAE, including changes
in the magnitude of the anisotropy barrier, as well as instances of
changes to an easy-plane system.  

To measure the direct effect of the electric field of the Br ions
on the spin-orbit coupling, external Coulomb potentials which acted
to cancel the long range affects of the Br anions were added and
the spin-orbit interaction and magnetic anisotropy were recomputed.
This neutralized the electric field due to the Br$^-$ ions near the
Mn sites without changing the electronic structure of the molecule.
We tested the effects with neutralizing charge distributions of
various widths, and by neutralizing four, eight, or all twelve Br
anions. The MAE changed by less than 1~K in all of these calculations.
We therefore conclude that the electric fields created by the Br$^-$
ions do not have a significant effect on the magnetic properties of
the molecule.

In conclusion, we present a study of the electronic and magnetic
properties of the Mn$_{10}$ single molecule magnet.  We confirm the
experimentally suggested magnetic ordering, although we find that
a state with $S=13$ is the magnetic ground state in contrast to the
$S=12$ state suggested from high field EPR measurements\cite{Mn10}.
In agreement with experiment we find the Mn$_{10}$ unit is an easy-axis
system with a small barrier of 9.5~K and
the compensating cluster in the molecular crystal, 
which has one Mn atom with $S=5/2$, is an easy plane system with a 
MAE of 0.1~K, negligible compared with the Mn$_{10}$ unit.
We show that the
magnetic anisotropy is determined by a competition between different
spin channels involved.  
The electric field caused by the negative charges of Br anions
has no significant direct effect on the spin-orbit coupling or the MAE,
although their chemical interactions do have significant effect on
the electronic structure and therefore on the MAE.
The states most important for the magnetic
anisotropy energy involve transitions between occupied majority and
unoccupied minority $d$-states at the same Mn-atom.

J.K. would like to thank the Schloe{\ss }mann Foundation for
financial support and R. Sessoli for helpful discussions. 
The work was partially supported by the Office of Naval
Research and the DOD HPCMO.

\end{document}